\documentclass[letterpaper]{JHEP3}

\usepackage{graphicx}

\usepackage{amsmath}
\usepackage{amsfonts}
\usepackage{epsfig}

\newcommand{\beq}{\begin{equation}}
\newcommand{\eeq}{\end{equation}}

\newcommand{\psib}{{\overline{\psi}}}

\newcommand{\half}{\frac{1}{2}}
\newcommand{\s}{\sigma}
\newcommand{\myref}[1]{(\ref{#1})}
\newcommand{\gappeq}{\mathrel{\rlap {\raise.5ex\hbox{$>$}}
{\lower.5ex\hbox{$\sim$}}}}
\newcommand{\lappeq}{\mathrel{\rlap{\raise.5ex\hbox{$<$}}
{\lower.5ex\hbox{$\sim$}}}}
\newcommand{\ord}[1]{{{\cal O}(#1)}}

\title{Phase diagram of $SU(2)$ with 2 flavors of dynamical adjoint quarks}

\author{Simon Catterall,
Department of Physics, Syracuse University, Syracuse, NY 13244, USA
E-mail: \email{smc@phy.syr.edu}
}
\author{Joel Giedt, Department of Physics, Applied Physics
and Astronomy, Rensselaer Polytechnic Institute,
Troy, NY 12180 USA E-mail: \email{giedtj@rpi.edu}}

\author{Francesco Sannino, HEP Center, Institute for Physics and Chemistry, University of Southern
Denmark, Odense, Denmark. 
E-mail: \email{sannino@ifk.sdu.edu}}
\author{Joe Schneible, Department of Physics, Syracuse University, NY 13244.
E-mail: \email{jschneib@physics.syr.edu}}

\preprint{}

\date{July 2008}

\abstract{
We report on numerical simulations of $SU(2)$ lattice gauge theory with
two flavors of light dynamical quarks in the adjoint of the gauge group.
The dynamics of this theory is thought to be very different from QCD --
the theory exhibiting conformal or near conformal behavior in the infrared.
We make a high resolution survey of the phase diagram of this model
in the plane of the bare coupling and quark mass on lattices
of size $8^3\times 16$. Our simulations reveal
a line of first order phase transitions extending from $\beta=0$ to
$\beta=\beta_c\sim
2.0$. For $\beta>\beta_c$ the phase boundary is no longer first order but continues
as the locus of minimum meson mass. For $\beta>\beta_c$ we
observe the pion and rho masses along the phase boundary
to be light, independent of
bare coupling 
and approximately degenerate. We discuss possible interpretations of
these observations and corresponding continuum limits.
}


\begin{document}

%
\section{Introduction}

Non-abelian gauge theories at zero temperature and matter density 
can exist in a number of distinct phases 
which can be distinguished by the characteristic dependence of
the potential energy on distance for
two well separated static sources.  These different behaviors of the
potential energy can be accessed by varying the number of colors and
the number of flavors of fermions.
The collection of all of these different behaviors,
when represented in the flavor-color space, 
constitutes the {\it Phase Diagram} of the given gauge theory. 
Up to possible dualities among different theories it uniquely
defines each theory.  In \cite{Sannino:2008ha} the reader will
find an up to date review of all of the possible phases for a generic gauge theory.

Knowing the phase diagram of strongly coupled theories has an
immediate impact on the construction of sensible extensions of
the standard model of particle interactions. Dynamical breaking
of the electroweak symmetry is a time-honored example. It is well
known that scaled up versions of QCD \cite{TC} are ruled out by
electroweak precision data.\footnote{The reader will find 
in \cite{Sannino:2008ha,Foadi:2007se} an exhaustive review of 
all of the precision data results from LEP I and II and 
how they constrain old and new models of dynamical breaking of the electroweak theory.}  

Using fermions in higher dimensional representations of the
gauge group opens up many new phenomenological
possibilities \cite{Sannino:2004qp,Dietrich:2005jn,Dietrich:2006cm}. 
There are, in fact, a number of reasons to recommend using higher dimensional representations in the underlying dynamics breaking the electroweak theory: i) The dynamics is generally different from QCD; ii) A near conformal behavior can be 
reached for a very low number of fermions naturally reducing the 
contribution to precision observables \cite{Sannino:2004qp}; iii) The spectrum of spin one states of these theories leads to interesting physical processes to be observed at the LHC \cite{Appelquist:1998xf,Appelquist:1999dq,Foadi:2007ue}.

An explicit phenomenological realization of this type of model 
is termed Minimal Walking Technicolor (MWT) \cite{Foadi:2007ue}
and is based on an $SU(2)$ gauge theory coupled to
two flavors of adjoint quarks. This model is thought to
lie close in theory space to theories with non-trivial infrared fixed points
\cite{Sannino:2004qp,Ryttov:2007cx}. Indeed it is possible that this
theory already exhibits such a fixed point.
In the vicinity of such a zero of the beta-function
the coupling constant flows slowly
or {\it walks}.
Originally such models
were introduced to alleviate the flavor changing neutral current problem for extensions of the technicolor theory needed to give mass to the standard model fermions \cite{Eichten:1979ah,Holdom:1981rm,Yamawaki:1985zg,Appelquist:an}. The MWT is thought to achieve such walking behavior with a minimal
number of light (techni)quarks \cite{Sannino:2004qp}. This is the theory 
studied numerically in this paper.

Another recent extension of the standard model which has attracted a great deal of interest is unparticle physics \cite{Georgi:2007ek}. One simply couples a new conformal sector to the standard model. It is natural to identify this sector  with a strongly coupled theory featuring an infrared fixed point.  Knowledge of the phase diagram is then essential to provide natural ultraviolet completions of unparticle models. Making use of the analytic knowledge of the phase diagram one finds, for example, that it is not easy to construct gauge theories with an infrared fixed point able to produce spinor-type ``unparticle stuff'' \cite{Sannino:2008ha}.

It is hence crucial to gain information on the phase 
diagram of strongly interacting gauge theories.  Lattice 
methods and computational resources are now mature enough
to provide a ``first principles'' systematic study of 
such phase diagrams, with dynamical fermions in
the chiral limit on reasonably large lattices.  Investigations
of representations other than the fundamental have just
begun \cite{Catterall:2007yx,DelDebbio:2008wb,Shamir:2008pb,
DelDebbio:2008zf}, significantly extending older work
on very small lattices \cite{kogut}.  Simulation studies
of many flavors in the fundamental representation have also become more
active of late \cite{Appelquist:2007hu,Deuzeman:2008sc}, 
extending the results of \cite{mawhinney,Iwasaki:2003de}.

In the current work we
provide a high resolution scan of the mass/coupling phase diagram (not to be
confused with colors/flavors/representation discussed above)
of $SU(2)$ gauge
theory with two (Dirac) flavors of fermions in the adjoint (triplet) representation,
using larger lattices and higher statistics than were utilized in our
earlier work \cite{Catterall:2007yx}.
We find clear evidence of a phase boundary in the two dimensional plane
of bare gauge coupling and quark mass. For $\beta<\beta_c\sim 2.0$ the system
undergoes a first order phase transition as this line is crossed. The latent
heat of this transition goes to zero for $\beta\to \beta_c$ while the
line continues to larger $\beta$ as the locus of minimum meson mass.
For $\beta < \beta_c$ we see evidence for chiral symmetry breaking and a Goldstone
behavior of the pion. Conversely, for $\beta > \beta_c$ 
the Goldstone behavior $m_\pi^2 \propto m_q$ disappears in a
a novel way as the system is tuned close to the phase boundary, the string tension
in lattice units is so small that we can only bound it
from above, and the pion and rho masses drop quickly
to values that are degenerate, within statistical
errors.  

All this behavior is indicative of a theory that
is very different from QCD, or even the theories with fundamental
flavors that have been studied on the lattice.  We will describe the
various possible interpretations of the lattice data 
in the Discussion section at the end
of this work, but here we would like
to highlight the most interesting one:  all of the lattice
data that we have obtained is consistent with either a nontrivial
infrared fixed point where the theory becomes conformal, or
a theory with an asymptotically free beta function that
is very small for the range of scales accessible on our
lattice (i.e., the coupling runs so slowly that we cannot
start at weak coupling and still have access to the confinement
scale on lattice of modest size).

Either way, if this theory were to provide the mechanism
of dynamical electroweak symmetry breaking, the phenomenology
would be radically different from a QCD-like technicolor scenario,
and most likely naive dimensional analysis arguments
would not be valid.  Thus, the lattice results that we have
obtained are quite exciting from this perspective
and warrant further studies on larger lattices, which
are currently in progress.

In the next section we summarize some of the relevant analytical
results.  We then describe our lattice model and present our numerical
results.  Finally, we end with a discussion and interpretation
of what we have found from our Monte Carlo study.  A full tabulation
of the meson masses that we have obtained is presented in Appendices A and B.

\section{Summary of the Analytical Results}
Dynamical fermion lattice simulations 
of higher dimensional representations 
are at an exploratory stage and it is hence useful to 
compare the results with theoretical expectations 
obtained using various analytical methods.   
To gain insight one can now use, for instance, the conjectured 
all-order beta function for nonsupersymmetric gauge 
theories \cite{Ryttov:2007cx} together with the 
constraints from the unitarity of the conformal operators.
This method constitutes a step forward with respect to 
the older approach based on the truncated Schwinger-Dyson
equation (SD) \cite{Appelquist:1988yc,Cohen:1988sq,Miransky:1996pd}, 
which is also referred to as the ``ladder approximation'' in the 
literature.  In contrast to the ladder approximation,
the all-order beta function allows one to determine 
the fermion mass anomalous dimension 
for any strongly coupled gauge theory at the infrared fixed point. 
Anomalous dimensions at fixed points are 
scheme independent since they represent physical quantities.
The analytical phase diagram obtained by this approach,
and a comparison of it to recent lattice results \cite{Catterall:2007yx,
Shamir:2008pb,DelDebbio:2008zf, Appelquist:2007hu,Deuzeman:2008sc},
is summarized in \cite{Sannino:2008ha}. 

In the ladder approximation the $SU(2)$ theory with two Dirac
flavors of adjoint fermions should be {\it just} 
below the conformal window where the theory develops 
an infrared fixed point~\cite{Sannino:2004qp}.
In the context of this approximation this means that the 
anomalous dimension of the fermion mass 
exceeds unity.  However, according to the 
all-order beta function, if the infrared fixed point is actually reached
then the anomalous dimension assumes the value
$\gamma = 3/4,$ where
$\gamma=-{{\rm d}\ln m}/{{\rm d}\ln \mu}$ and $m$ is 
the running fermion mass.  
If we take $\gamma=1$ as the boundary of
the conformal window, the all-order beta function 
suggests that the $SU(2)$ model is conformal in the infrared.
However, the constraint coming from the unitarity allows $\gamma$ to be as
large as two before conformality is lost.  Thus it
is an open question whether or not a nontrivial infrared
fixed point exists.  As will be seen, the results of
our lattice study suggest that this fixed point may
exist, though further investigations will be required
to strengthen the case for that conclusion.

It is instructive to compare this theory with the case of 
the $SU(3)$ gauge theory with two Dirac fermions in the two
index symmetric representation.  In ``theory space'' the 
previous gauge theory and the present one are very close, 
since the adjoint of $SU(2)$ is equivalent to 
the two index symmetric representation.  Recent lattice 
results \cite{Shamir:2008pb} suggest that this 
theory may have an infrared fixed point, though 
more studies are needed here too.  We note that the ladder approximation
predicts that this theory is nearly conformal ({\it i.e.} walking),
and further away from conformality then the $SU(2)$ theory.
Also, if one assumes that the theory is conformal 
in the infrared, then the all-order beta function predicts that
the anomalous dimension of the fermion condensate is $\gamma=1.3$,
larger than the value of $3/4$ that was found in the SU(2) case
above.  If it is true that $SU(3)$ has an infrared fixed point,
it follows that the $SU(2)$ theory also has an infrared fixed point,
since the screening due to fermions is even greater in the latter case.

As an aside, we note that it is quite interesting that for $SU(3)$
the anomalous dimension $\gamma$ is larger than unity.  If true, this would 
be quite an important result, since large anomalous
dimensions are needed when constructing extended 
technicolor models that are able to account for 
the heavy quark masses, as noted in \cite{Sannino:2008ha}.
If the preliminary indications of $\gamma>1$ hold up
to further scrutiny, it would overturn
the common lore---but no rigorous theorem---regarding the anomalous dimension 
of the ``quark'' bilinear operator.

Other interesting cases to consider are those with eight and twelve 
Dirac fermions in the fundamental representation of SU(3). 
The all-order beta function predicts that the conformal window
cannot be achieved for a number of flavors less then 8.25 (really,
nine once the integer constraint is imposed) for
the fundamental representation of $SU(3)$.  This is confirmed
by the latest lattice results \cite{Appelquist:2007hu,Deuzeman:2008sc}. 
In that work it was also suggested that the theory 
with twelve flavors has an infrared fixed point. 
The prediction of the anomalous dimension of the 
quark mass operator is then $\gamma = \frac{3}{4}$.
Amusingly this theory has the same anomalous dimension
as the $SU(2)$ two adjoint flavor theory that we study
here (assuming they both possess an infrared fixed point).

\section{Lattice Implementation}
\subsection{Action and simulation algorithm}
The lattice action we employ consists of the usual Wilson plaquette term 
\begin{equation}
S_G=-\frac{\beta}{2}\sum_x\sum_{\mu>\nu}{\rm Re}{\rm Tr}\left(
U_\mu(x)U_\nu(x+\hat\mu)U^\dagger_\mu(x+\hat\nu)U^\dagger_\nu(x)\right) \ ,
\end{equation}
with the link matrices $U_\mu(x)$ in the fundamental representation of
SU(2), together with the Wilson action for two Dirac fermions in the adjoint
representation:
\begin{eqnarray}
S_F&=&-\frac{1}{2}
\sum_x\sum_\mu\psib(x)\left(
V_\mu(x)\left(I-\gamma_\mu\right)\psi(x+\mu)+
V^T_\mu(x-\mu)\left(I+\gamma_\mu\right)\psi(x-\mu)
\right)\\
&+&\sum_x \left(m+4\right)\sum_x\psib(x)\psi(x) \ .
\end{eqnarray}
Here adjoint links $V_\mu(x)$ are used, which are related to
the fundamental links by
\begin{equation}
V^{ab}_\mu(x)=\half {\rm Tr} \left( \s^a U_\mu(x) \s^b U^\dagger_\mu(x) \right) \ ,
\end{equation}
with $\s^a, \, a=1,2,3$ the usual Pauli matrices.

We have simulated this theory over a range of
gauge couplings $\beta=1.5 - 3.0$ and bare quark masses 
$m$ ranging from $ -2.0 < m <0.5 $ on $8^3\times 16$ lattices 
using the usual Hybrid Monte Carlo algorithm
\cite{hmc}. 
Typically we have generated between $400-2000$
$\tau=1$ HMC trajectories.  Antiperiodic boundary conditions were used for
the fermions in the time direction (in order to ameliorate
problems with exceptional configurations at the
for small quark mass), whereas all other boundary conditions
are periodic.

All simuluations were run on the IBM BlueGene/L SUR machine
at Rensselaer over a period of four months.  The simulation
software used is a recent, BlueGene/L architecture-specific
version of the Columbia Physics System, modified such that 
SU(2) with any number of adjoint (Wilson or domain wall) fermions can be studied.
The code has been validated by reproducing the results
of \cite{Fleming:2000fa} for the case of pure super-Yang-Mills.
Indeed, the software was developed for a large-scale follow-up
study of pure super-Yang-Mills that is in progress~\cite{Gie08}.
The average compute rate was 70 Gflop/s, on a 128 node partition
of the BlueGene/L.

\subsection{Meson operators}
We estimate the hadron masses by suitable fits to
corresponding time sliced averaged
correlation functions
\begin{equation}
G_O(t)=\sum_{x,y}<\psib(x,t)\Gamma_O\psi(x,t)\psib(y,0)\Gamma_O\psi(y,0)>
 \,\end{equation}
where
$\Gamma_O=\gamma_5$ for the pion and $\Gamma_O=\gamma_\mu, \mu=1,2,3$ for
the rho (the latter being averaged over spatial directions $\mu$).
Errors are estimated by a jackknife procedure in which fits are made to
the meson correlators using subsets of the data, the mean and deviation
of the resulting mass distribution yielding a mean meson mass and
error. 

\subsection{String tension}
We estimate the string tension as a function of lattice scale $R$ from the
large distance asymptotic behavior of the Creutz ratio \cite{Creutz:1980wj}
\beq
\chi(R,R) = -\ln{\frac{W(R,R)W(R-1,R-1)}{W(R,R-1)W(R-1,R)}} \sim \sigma a^2 .
\label{ceq}
\eeq
Here $W(R,R')$ is the expectation value of the $R \times R'$ Wilson loop
and $a$ is the lattice spacing.  The asymptotic behavior on the r.h.s.~of
\myref{ceq} assumes an area law for the Wilson loops.  In practice
one looks for the Creutz ratios $\chi(R,R)$ to coalesce on an envelope
where the area law becomes dominant.  This occurs for $Ra$ of order
or larger than the scale of confinement $\ell_c = 1/\Lambda$,
where $\Lambda$ is the usual dynamical scale of an asymptotically
free gauge theory.  In the chiral limit where
the fermions are massless, the lattice spacing
is a function of the bare lattice coupling $\beta=4/g^2$, through
$a \sim \Lambda^{-1} \exp(-2 \pi^2 \beta /b)$ where $b$ is the 1-loop
beta function coefficient.  (Of course this
estimate can be improved with high loop results, as
has been considered in \cite{DelDebbio:2008wb}.)  
Thus since $\s$ and $\Lambda$ are physical
scales, one expects to see an exponentially decreasing envelope
for the Creutz ratios.  Away from the chiral limit, there is
a threshold mass above which the running of the coupling is altered.
In that case one would have $a=a(\beta,m)$, where $m$ is the
bare fermion mass.  Finally, we should note that since we use
fundamental links in the Wilson loops, they are not screened by
the adjoint fermions, and an area law emerges at 
scales $Ra > \ell_c$, provided the theory is asymptotically free.
On the other hand, if the theory has a nontrivial infrared fixed
point, the only scale available is the finite extent of the lattice,
$L^3 \times T$ (here, dimensionless quantities).  One would therefore expect to see that the
Creutz ratio behavior depends strongly on $L,T$, in contrast
to what happens in the confining case where for $\sigma a^2 L^2$
and $\sigma a^2 T^2$ very large the results become independent
of $L,T$.

\section{Results}
\begin{figure}
\begin{center}
\includegraphics[width=4.5in,height=3.2in,bb=0 50 800 550,clip]{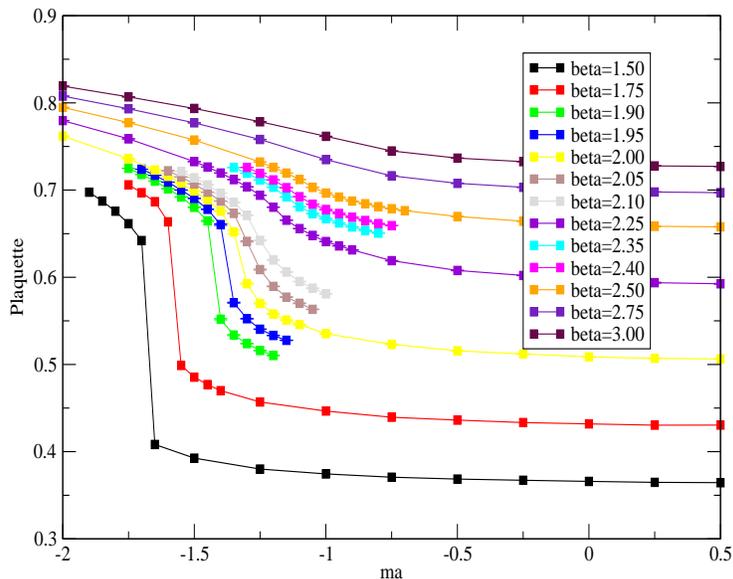}
\caption{Plaquette expectation values as a function of the bare
fermion mass, along lines of constant lattice gauge coupling $\beta=4/g^2$.
It can be seen that $\beta_c \approx 2$ marks a transition, below which
a first order phase transition is seen as the
quark mass is varied.  We therefore find
that close to the phase boundary, $\beta_c$ corresponds
corresponds to a ``bulk'' transition, below which only a lattice
phase exists.  This can be understood in terms of the dynamical
generation of an effective adjoint plaquette term in the gauge
action, due to the radiative effects of nearly massless adjoint
``quarks.''  Of course, for masses far enough away from the critical
value the renormalization of the gauge action is relatively small
and the adjoint term will not lead to a bulk transition.}
\label{fig1}
\end{center}
\end{figure}

\begin{figure}
\begin{center}
\includegraphics[width=4.5in,height=3.2in,bb=0 50 800 550,clip]{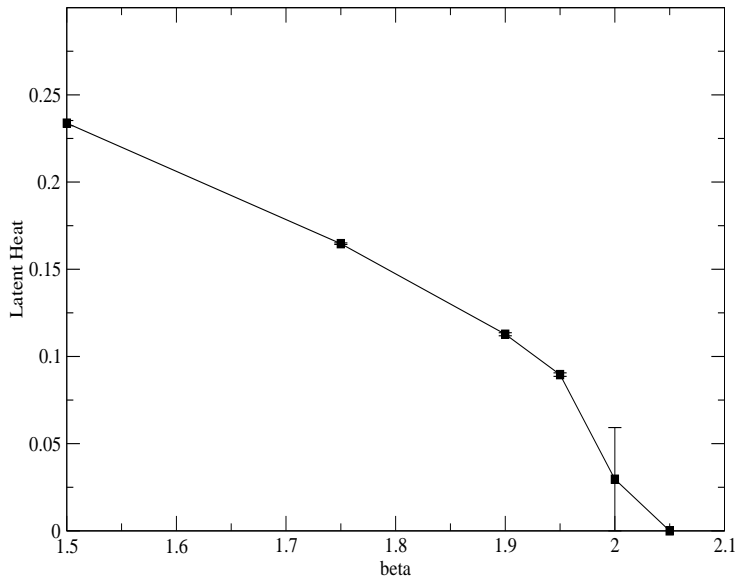}
\caption{The latent heat, which appears to vanish in the
$\beta \to 2$ limit.}
\label{fig2}
\end{center}
\end{figure}

Our results were obtained as a series of bare Wilson fermion mass scans,
at fixed bare gauge coupling $\beta=4/g^2$.  
Perhaps the simplest observable to consider is
the expectation value of the plaquette or action. 
Fig.~\ref{fig1} shows a series
of scans in the ``quark'' mass for different values of $\beta$.  (In
what follows we will often refer to the elementary fermions as ``quarks''
and composite states as ``pions'', ``rhos'', etc.  Of course this
is only by way of analogy, and we could alternatively prefix these
names with ``techni-''.)
Notice the appearance of a discontinuity for small $\beta< \beta_c \sim 2.0$.
The data indicates that a line of first order phase transitions exists for small
$\beta$.  Further support for this conclusion comes from
the latent heat, as measured by the jump in the plaquette
and displayed in Fig.~\ref{fig2}.  It appears to vanish as $\beta \to
\beta_c\sim 2.0$.
Beyond $\beta_c$ we observe that the phase boundary continues as the locus
of minimum pion and rho meson mass.  The natural conclusion is that
$\beta_c \approx 2$ is a second order end-point for the line of first
order transitions.  In Section \ref{disc} we will interpret the first
order behavior across the phase boundary, at $\beta < \beta_c$,
as corresponding to a bulk
phase transition in the effective SU(2) gauge theory, in accordance
with the well-known combined fundamental/adjoint plaquette action
phase diagram \cite{Greensite:1981hw,Bhanot:1981eb}.

\begin{figure}
\begin{center}
\includegraphics[width=4.5in,height=3.2in,bb=0 50 800 550,clip]{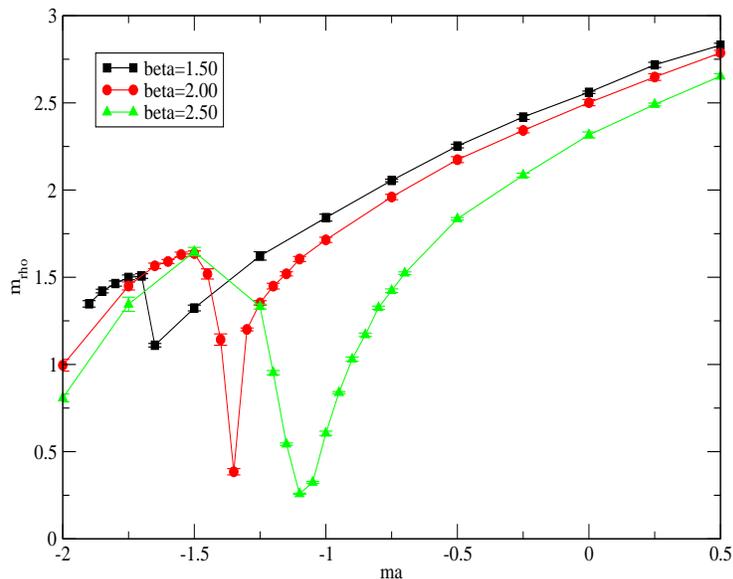}
\caption{The ``rho'' mass $m_\rho a$,
as a function of the bare Wilson fermion
mass $m$, for three example values of the bare lattice coupling $\beta$.
Note that as $\beta$ increases past the critical value $\beta_c \approx 2$,
the $\rho$ mass on the phase boundary becomes small on the
order of the inverse lattice size $1/L$.
This is consistent with
the $\rho$ becoming a massless state in the thermodynamic limit}
\label{fig3}
\end{center}
\end{figure}

In Fig.~\ref{fig3} we illustrate the behavior of the rho
mass $m_\rho a$ by plotting it as a function of the bare Wilson fermion mass $ma$
at three representative
points in the phase diagram:  $\beta=1.5$, $\beta=2.0$ and $\beta=2.5$. 
(The full set of rho mass results from our studies is tabulated
in Appendix B.)
The region of quark mass to the left of the
minimum corresponds to an Aoki phase \cite{Aoki}, except that
here it is for adjoint Wilson fermions, which was
recently studied in~\cite{DelDebbio:2008wb}\footnote{Notice that the
dependence of the rho mass on bare quark mass appears to depart from
linear at strong coupling which we attribute to the proximity of the
first order phase transition}
.
In the case of $\beta \geq 2$, as one approaches the minimum meson mass
from above, there is a rapid drop in $m_{\rho}$ that is inconsistent with
a simple linear variation with bare quark mass $m_\rho \propto m-m_c$,
as one would see in QCD, or in the case of two fundamental
flavors observed in \cite{Catterall:2007yx} (cf.~Fig.~5 of that
reference).  The data corresponds instead to a form
\beq 
m_{\rho} \sim (m-m_c)^{1/(1-\epsilon)}, \quad 0<\epsilon<1.
\eeq

Similar results are obtained for the pion, illustrated
for same three values of $\beta$ in Fig.~\ref{fig4} below.
(A full tabulation of pion masses is given in Appendix A.)
The figure shows the $(m_\pi a)^2$ as a function of
the bare quark mass $ma$, so it is important to keep in mind
that the dependence of the lattice spacing on $\beta,m$ also
enters into the plot.
There is clear evidence of a linear Goldstone dependence at strong
coupling consistent with chiral symmetry breaking for $\beta<2.0$. Conversely
at $\beta=2.5$ the data near the phase boundary line is consistent with a simple linear
dependence of the pion mass on bare quark mass and chiral symmetry 
restoration.  Again, the
pion mass varies very rapidly with bare quark mass close to $\beta=2.0$.
One interpretation, which we will discuss further below,
is that the lattice spacing shrinks significantly as one
approaches the chiral limit, due to the comparable renormalization
of the gauge coupling by gluons and and quarks in this model.  I.e.,
the coupling walks when the quarks are very light, and does
so over a large range of scales as the quarks approach zero 
renormalized mass.

\begin{figure}
\begin{center}
\includegraphics[width=4.5in,height=3.2in,bb=0 50 800 550,clip]{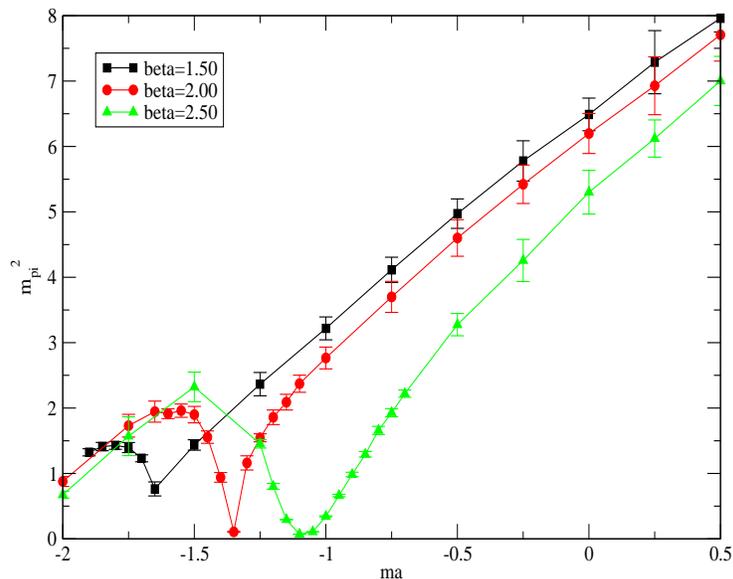}
\caption{The pion mass squared for example values of $\beta$.
The very sharp behavior as the bare mass is varied away from the
phase boundary near
$\beta=2$ stands in contrast to the rounding that
would normally be expected from the effects of the finite size effects.
The significant decrease in the slope of the line as
one approaches the phase boundary is presumably due to
$(m_\pi a)^2/(m a) \sim a$, with $a(\beta,m)$ having
a significant $m$ dependence when the fermions are very light.
This is particularly true since the contribution of the quarks to
the running of the coupling is quite close to that of the gluons.}
\label{fig4}
\end{center}
\end{figure}

\begin{figure}
\begin{center}
\includegraphics[width=4.5in,height=3.2in,bb=0 50 800 550,clip]{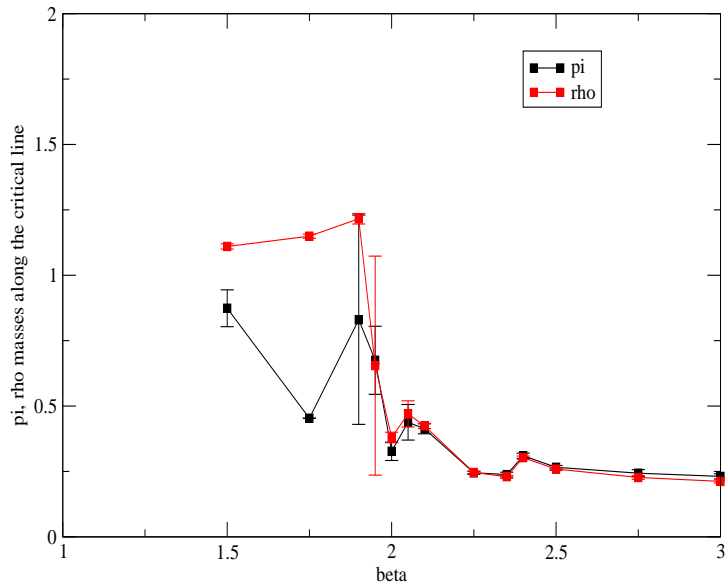}
\caption{Pion and rho masses along the phase boundary.  Note that
they become degenerate for $\beta \gappeq 2$.}
\label{fig5}
\end{center}
\end{figure}

The behavior of the pion and rho masses along the entire phase boundary
is shown in
Fig.~\ref{fig5}. Two regimes are seen; a strong coupling phase with a light
pion and heavy rho for $\beta < \beta_c \sim 2.0$ and a phase for
$\beta>\beta_c$ where the pion and rho
are approximately degenerate and independent of the bare coupling. The situation
at $\beta\sim\beta_c$ is somewhat unclear as the statistical errors are large
there.  The phase boundary itself is shown below in Fig.~\ref{fig6}.

\begin{figure}
\begin{center}
\includegraphics[width=4.5in,height=3.2in,bb=0 50 800 550,clip]{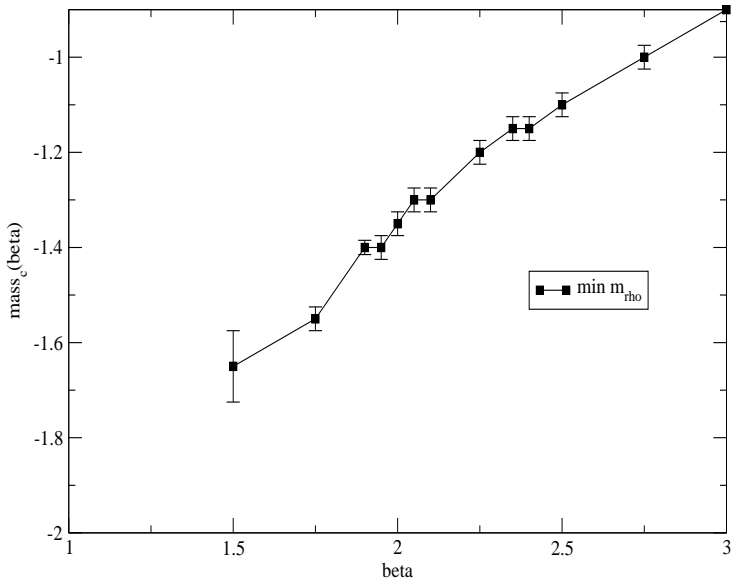}
\caption{Here, the phase boundary is extracted from the 
minimum of rho mass at each $\beta$.}
\label{fig6}
\end{center}
\end{figure}

\begin{figure}
\begin{center}
\includegraphics[width=4.5in,height=3.2in,bb=0 50 800 550,clip]{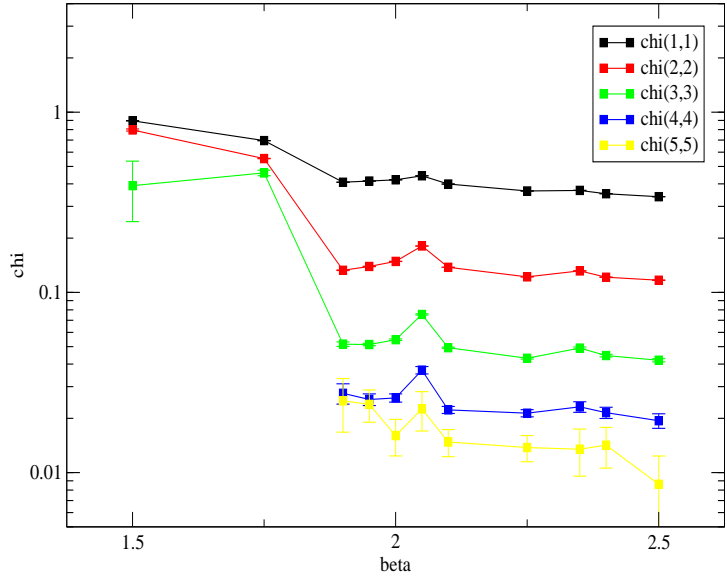}
\caption{Here we show the Creutz ratios along the phase boundary corresponding
to minimum meson masses.
For $\beta=1.9, 1.95$ the envelope that determines $\sigma a^2$
can be seen from  $\chi(4,4)$ and $\chi(5,5)$, since they coincide.  For $\beta \geq 2$
the envelope where $\chi$'s begin to coalesce cannot be seen, though
the value of $\chi(5,5)$ places an upper bound on $\sigma a^2$.
For $\beta < 2$ there is some indication of an envelope, though large statistical errors
prevent us from measuring the larger loops needed for $\chi(4,4)$
and $\chi(5,5)$.}
\label{fig7}
\end{center}
\end{figure}

We have also made estimates of the string tension
as measured by Creutz ratios $\chi(R,R)$ [cf.~\myref{ceq}] 
of various sizes.  Fig~\ref{fig7} shows a plot
of $\chi(R,R)$, $R=1,\ldots,5$, as a function of 
$\beta$ as we move along the phase boundary $m_c(\beta)$.
At distances of order or larger than the confinement scale, these
ratios should coalesce on the value of $\sigma a^2$ where $\sigma$
is the string tension.  For $\beta=1.9, 1.95$ this occurs, as it
can be seen that $\chi(4,4)$ and $\chi(5,5)$ coincide.  For $\beta \geq 2$
the envelope where $\chi$'s begin to converge cannot be seen, but
the value of $\chi(5,5)$ places an upper bound on $\sigma a^2$.  It may
be that much larger $R$ values in $\chi(R,R)$ are needed, which is
not possible on the $8^3 \times 16$ lattice that we study here.
This would be the case if the lattice spacing $a$ has the very sensitive exponential
dependence on $\beta$ that would be expected from a walking theory;
for instance in the present theory using 2-loop running one would predict that
between $\beta=2$ and $\beta=2.1$, $\sigma a^2$ would decrease by an
order of magnitude and between $\beta=2$ and $\beta=2.5$ it would decrease
by five orders of magnitude.  Given the trend in Creutz ratios with
$R$ at $\beta=2$, one can roughly estimate that $\chi(7,7)$ or $\chi(8,8)$
may be required before the envelope at this value of $\beta$ would be
seen.  This would require a lattice of size $16^3 \times 32$, which
we are currently studying.
On the other hand, and this is the possibility
that we would like to emphasize, it could be that for $\beta \geq 2$ one falls
into the basin of attraction for a nontrivial infrared fixed point,
and the area law does not hold at any scale.  

For $\beta < 2$ one
has hints of the envelope, though the large statistical errors due
to enhanced fluctuations at small $\beta$
prevent us from measuring the larger loops needed for $\chi(4,4)$
and $\chi(5,5)$.  Nevertheless, it would appear that the string tension $\s$
is of order $1/a^2$, consistent with a phase of the theory
dominated by lattice artifacts.

In the next section we discuss further the
possible interpretations of these observations.

\section{Discussion}
\label{disc}
Generically a lattice gauge theory will have a confining phase at
strong bare coupling, and we believe this to be true for the
present theory.  Typically this is signaled by
a non-zero string tension extracted from the asymptotic
behavior of Wilson loops, or correlation functions of
Wilson/Polyakov lines.
However at strong coupling the lattice theory
will be dominated by lattice artifacts.  For instance,
from Fig.~\ref{fig7} on sees that at $\beta < \beta_c \approx 2$,
the Creutz ratios indicate that $\s a^2 = \ord{1}$, so that
for small $\beta$ (strong coupling)
the string tension,
and hence scale of confinement, is of the same scale as
the lattice spacing $a$.  Similarly, in Fig.~\ref{fig3} on
sees that for $\beta=1.5$ we obtain $m_\rho a \geq 1$,
indicating that the rho also lies at the ultraviolet cutoff scale.

To understand whether this confining strongly coupled
phase survives the continuum limit it is necessary to examine the behavior
of, say, a Wilson loop, as the lattice spacing is sent
to zero holding the area of the Wilson loop fixed in physical units. 
For a theory exhibiting asymptotic freedom
this is accomplished by increasing $\beta$. 
In the case of QCD and on an infinite
lattice this process can
be continued indefinitely until we end up at the fixed point $\beta=\infty$,
thereby removing the ultraviolet cutoff.
However it is possible that this procedure
is interrupted by the presence of a first order phase transition at some
finite bare coupling---so that the signal of confinement at
strong coupling is not a property of
the continuum theory.  This appears to happen in this model,
in the vicinity of the first order line.  That is, our
results indicate that 
the strong coupling phase 
at $\beta < \beta_c$ is not continuously
connected to a phase with a continuum limit.

On the other hand, it does appear that one can move
smoothly into a continuum phase if one starts sufficiently
far away from phase boundary.  In fact, this must be
true since for large enough mass the theory is an arbitrarily
good approximation of the quenched theory with just the Wilson
plaquette action in the fundamental representation.  Since that
theory does not have a discontinuity separating the strong and
weak coupling phases, we know that this is also true in our
theory in this quenched limit.  

It is also possible that the a continuum theory with massless quarks
may be obtained by tuning the bare quark mass in the regime $\beta>\beta_c$.
The fact that the minimum meson mass appears to scale with the inverse
lattice size is consistent with this. 

All of this can be
understood in terms of radiative effects of two flavors of adjoint fermions.  
If the fermions are very
light, they will generate a large
adjoint plaquette term  when they are integrated out to
obtain the long wavelength effective theory.
As mentioned briefly above, it is known that a first order transition 
occurs in the
adjoint plaquette action theory \cite{Greensite:1981hw} and that in the
mixed fundamental/adjoint plaquette theory there is a
first order transition line when the adjoint term is
sufficiently large \cite{Bhanot:1981eb}.  The interpretation of our
results is therefore clear:  if the quarks are approximately
massless, the
adjoint fermions lead to a large effective adjoint plaquette
term, conventionally characterized by the
coefficient $\beta_A$.  As one passes through the phase boundary
at small fundamental
plaquette action coupling $\beta$, one is actually moving
back and forth across the first order line in the $\beta$--$\beta_A$
plane.  Since in the quenched fundamental/adjoint theory the
first order line only exists for $\beta<1.6, \beta_A > 0.7$, we can understand
why we too see that the first order behavior disappears for sufficiently
large $\beta$ or mass $m$.  The fact that the transition in our theory happens
instead at $\beta_c \approx 2$ would again be due to the radiative effects of
the fermions, which will also renormalize $\beta$.

Since in a walking technicolor scenario
we are interested in the theory with massless fermions,
the phase of the lattice theory that is relevant for continuum physics is the
phase where $\beta > \beta_c$.  For $\beta \to \infty$ we expect 
that the theory is
driven to the asymptotically free fixed point known to exist in perturbation
theory.  However the behavior of the theory in the infrared is less clear.
If the theory admits a new conformal fixed point then one expects that
this will govern the long distance physics of the model and long distance
features of the theory will be insensitive to the bare lattice coupling.
In addition such a theory has no intrinsic scale, so that the
only scales would be the lattice volume and the temperature.
It follows that both the string tension and
meson masses would scale to zero in the zero temperature,
thermodynamic limit.  This sort of behavior is certainly
consistent with what we see in the phase $\beta > \beta_c$
along the line of minumum meson mass $m=m_c(\beta)$.  Thus, our findings
are {\it consistent with the appearance of a new conformal fixed point
in this theory,} though they also leave open the possibility
of a walking theory.

However, one must be careful in drawing the conclusion that
a nontrivial infrared fixed point exists.  To take the
continuum limit along the critical line requires tuning the bare coupling
$\beta$ with lattice spacing $a$ such that finite
size effects are under control.  If this running $\beta(a)$ is sufficiently
slow it can lead to extreme sensitivity in the dependence of the lattice
spacing on bare coupling, when inverted to give $a=a(\beta)$.
As was discussed in relation to the Creutz ratio data above,
small increases in the $\beta$ would yield huge decreases 
in the lattice spacing.  It is difficult to analyze such
changes of scale on a relatively small lattice.  In particular,
large finite size effects can
mask the true infinite volume, zero temperature physics.  For example, the physical box size
can become so small that the system deconfines and looks quasi-free,
which would also be consistent with our data.
In effect, the physics is indeed being dominated 
by a conformal fixed point---not a new infrared stable 
point but the usual infrared unstable asymptotically
free fixed point.  To distinguish amongst the possibilites
will require larger lattices and a thorough study of finite size effects,
so one must be cautious in interpreting our findings thus far.

Another way of restating this is that any theory whose coupling runs
very slowly with scale will necessarily generate 
a dynamical mass scale in lattice
units (e.g., $a \Lambda$), which is very small for a weak bare coupling. 
To distinguish a confining theory
with a small scale from a theory with a non-trivial infrared fixed
point will then necessitate simulations on lattices which are significantly
bigger, in lattice units, than the inverse of this small mass scale, which is
a hard problem.  And, indeed, on small lattices the physics will be
governed by the usual ultraviolet fixed point corresponding to asymptotic
freedom.  On the other hand, simulations on larger lattices would allow us
to perform a ``step-scaling'' analysis, in order
to extrapolate to the infinite lattice volume behavior.
We have begun studies of $16^3 \times 32$ lattices with
the purpose of distinguishing between the walking and conformal
scenarios that we have just described.

\begin{figure}
\begin{center}
\includegraphics[width=3.5in,height=3in]{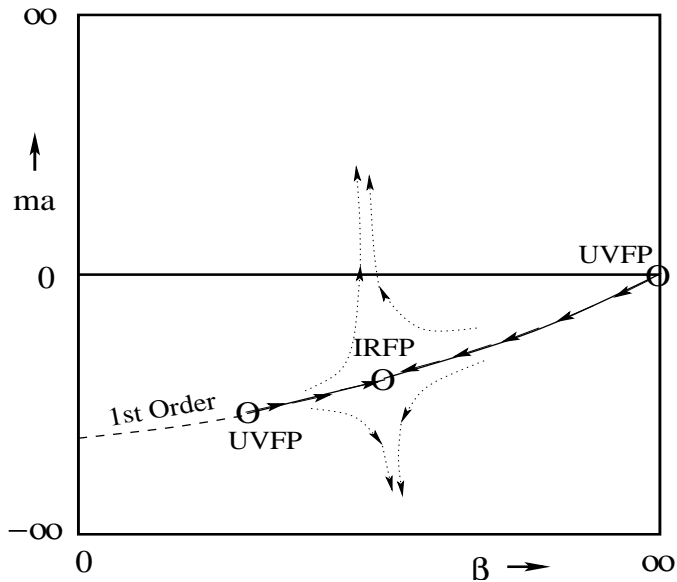}
\caption{
Here we provide a cartoon of renormalization group flows,
in order to illustrate what may occur if the present theory has
a nontrivial infrared fixed point.  The ``UV'' point on the right
corresponds to $(\beta,ma)=(\infty,0)$, the usual infrared unstable
fixed point.  On the other hand, ``CFP'' indicates a putative
nontrivial infrared fixed point where the theory is conformal.
Flows under a blocking transformation are indicated by arrows
along the line of critical masses.  Away from this line the
theory will flow to a quenched fixed point.  The infrared sink
is located at the far left-hand side, presumably $(\beta,ma)=(0,-2)$
based on Fig.~6.}
\label{fig8}
\end{center}
\end{figure}

Finally we would like to conclude by discussing a possible phase diagram which
might be relevant in the situation where the theory does indeed
contain a new conformal fixed point.  Fig.~\ref{fig8} shows a cartoon of
fixed points and possible RG flows for this model projected to the
plane of bare coupling constants $(\beta,m)$.  The arrows denote the
flow of couplings under {\it increases} in length scale corresponding, for
example, to a blocking transformation.
The theory certainly
contains the usual infrared unstable
fixed point corresponding to $(\beta,ma)=(\infty,0)$. A critical line 
corresponding to massless quarks extends out of this fixed point
to smaller $\beta$ or stronger coupling. If a conformal fixed point exists
it should form a sink for these flows as shown. 

In the picture we also show as a dashed line the line of first
order phase transitions. Our data supports the conjecture that this
line ends on a critical point corresponding to another infrared unstable
fixed point. 

Furthermore, our results are consistent with the first order line 
and the critical
line joining together at the critical endpoint. 
Any putative conformal fixed point
would then serve as a infrared sink
for massless flows out of these
fixed points as
shown\footnote{Many thanks to Ben Svetitsky for illuminating discussions of
these issues}.

Notice that all these fixed points are
also unstable in the direction orthogonal to the critical line; i.e. under
a mass deformation. Recognizing this fact actually allows us to
draw a RG flow that would automatically permit a {\it walking dynamics}
even in a theory inside the conformal window.  One merely allows the theory to
start near one of ultraviolet fixed points with a small but non-zero mass.
Under blocking such a trajectory would flow 
initially towards the conformal fixed point in the vicinity of which the flow
would slow before eventually flowing out along a direction corresponding to
a mass deformation.
Of course a walking scenario that introduces a mass for the fermions is not
what is desired when trying to use the theory studied here for breaking the
electroweak symmetry dynamically. Nevertheless, the possibility of a
nontrivial infrared fixed point in the present theory would be interesting
in its own right. However, there are technicolor models which make use of
different gauge dynamics realizing this possibility as explained in
\cite{Sannino:2008kg,Luty:2008vs}.

In conclusion, we have found that the present theory at
critical fermion mass either has a very slowly running gauge 
coupling (walking) or a nontrivial infrared fixed point (conformal).
The behavior is drastically different from theories that
do not sit near the conformal window.  However, special
difficulties emerge in the present theory due to extreme
sensitivity of the lattice spacing on the bare gauge coupling,
due to the slow running.  Studies on larger lattices and
a careful step-scaling analysis is needed, and indeed underway,
in order to clarify these issues.

\acknowledgments Catterall is supported in part by DOE grant
DE-FG02-85ER40237. The authors would like to acknowledge useful
conversations with George Fleming.  JG expresses appreciation
to Pavlos Vranas for providing a copy of the code that was used
in \cite{Fleming:2000fa}, which was used as a basis for the
modifications to the current version of the Columbia Physics
System.  At various points JG also benefited from technical
assistance provided by Chulwoo Jung (Brookhaven National Lab) and
Adam Todorski (SCOREC and CCNI at Rensselaer).  
The computational efforts on this project, which utilized
the SUR BlueGene/L at Rensselaer, were supported 
by NSF grant 0420703 entitled ``MRI: Acquisition of Infrastructure for Research 
in Grid Computing and Multiscale Systems Computation''
and a gift by the IBM Corporation of a BlueGene/L computer.
JG acknowledges support from Rensselaer faculty development funds.

\newpage

\section*{Appendix A:  Pion masses}

\begin{center}
\begin{tabular}{ccc}
\begin{tabular}{|ccc|} \hline
 m & pion & error \\ \hline 
0.50 & 2.821& 0.013 \\ 
0.25 & 2.699& 0.033\\ 
0.00 & 2.547 & 0.019 \\ 
-0.25 & 2.403 & 0.026 \\ 
-0.50 & 2.229 & 0.022 \\ 
-0.75 & 2.028 & 0.023 \\ 
-1.00 & 1.793 & 0.027\\ 
-1.25 & 1.538 & 0.037 \\ 
-1.50 & 1.197 & 0.027 \\ 
-1.65 & 0.873 & 0.070 \\ 
-1.70 & 1.109 & 0.021 \\ 
-1.75 & 1.181 & 0.027 \\ 
-1.80 & 1.195 & 0.020 \\ 
-1.85 & 1.187 & 0.022 \\ 
-1.90 & 1.150 & 0.020 \\ 
\hline
\multicolumn{3}{|c|}{\bf Pion masses for $\beta=1.50$.}\\ \hline
\end{tabular}
& \hspace{0.5in} &
\begin{tabular}{|ccc|} \hline
 m & pion & error \\ \hline 
0.50 & 2.802 & 0.019 \\ 
0.25 & 2.693 & 0.018 \\ 
0.00 & 2.527 & 0.029 \\ 
-0.25 & 2.370 & 0.029 \\ 
-0.50 & 2.200 & 0.032 \\ 
-0.75 & 1.999 & 0.024 \\ 
-1.00 & 1.744 & 0.041 \\ 
-1.25 & 1.481 & 0.033 \\ 
-1.40 & 1.242 & 0.012 \\ 
-1.45 & 1.161 & 0.017 \\ 
-1.50 & 1.061 & 0.015 \\ 
-1.55 & 0.951 & 0.014 \\ 
-1.60 & 0.711 & 0.048 \\ 
-1.65 & 1.304 & 0.029 \\ 
-1.70 & 1.312 & 0.017 \\ 
-1.75 & 1.291 & 0.020 \\ 
\hline
\multicolumn{3}{|c|}{\bf Pion masses for $\beta=1.75$.}\\ \hline
\end{tabular}
\end{tabular}
\end{center}

\begin{center}
\begin{tabular}{ccc}
\begin{tabular}{|ccc|} \hline
 m & pion & error \\ \hline 
-1.20 & 1.457 & 0.011 \\ 
-1.25 & 1.385 & 0.012 \\ 
-1.30 & 1.305 & 0.010 \\ 
-1.35 & 1.195 & 0.011 \\ 
-1.40 & 1.048 & 0.015 \\ 
-1.45 & 0.82 & 0.42 \\ 
-1.50 & 1.250 & 0.026 \\ 
-1.55 & 1.351 & 0.016 \\ 
-1.60 & 1.380 & 0.0075 \\ 
-1.65 & 1.386 & 0.016 \\ 
-1.70 & 1.360 & 0.012 \\ 
-1.75 & 1.3367 & 0.0079 \\ 
\hline
\multicolumn{3}{|c|}{\bf Pion masses for $\beta=1.90$.}\\ \hline
\end{tabular}
& \hspace{0.5in} &
\begin{tabular}{|ccc|} \hline
 m & pion & error \\ \hline 
-1.15 & 1.492 & 0.014 \\ 
-1.20 & 1.430 & 0.016 \\ 
-1.25 & 1.340 & 0.018 \\ 
-1.30 & 1.229 & 0.018 \\ 
-1.35 & 1.094 & 0.027 \\ 
-1.40 & 0.67 & 0.13 \\ 
-1.45 & 1.146 & 0.025 \\ 
-1.50 & 1.320 & 0.016 \\ 
-1.55 & 1.392 & 0.020 \\ 
-1.60 & 1.402 & 0.018 \\ 
-1.65 & 1.378 & 0.026 \\ 
-1.70 & 1.358 & 0.028 \\ 
\hline
\multicolumn{3}{|c|}{\bf Pion masses for $\beta=1.95$.}\\ \hline
\end{tabular}
\end{tabular}
\end{center}

\begin{center}
\begin{tabular}{ccc}
\begin{tabular}{|ccc|} \hline
 m & pion & error \\ \hline 
-1.05 & 1.550 & 0.015 \\ 
-1.10 & 1.466 & 0.018 \\ 
-1.15 & 1.384 & 0.012 \\ 
-1.20 & 1.249 & 0.025 \\ 
-1.25 & 1.041 & 0.028 \\ 
-1.30 & 0.437 & 0.068 \\ 
-1.35 & 0.693 & 0.014 \\ 
-1.40 & 1.119 & 0.017 \\ 
-1.45 & 1.322 & 0.020 \\ 
-1.50 & 1.409 & 0.027 \\ 
-1.55 & 1.416 & 0.032 \\ 
-1.60 & 1.428 & 0.012 \\ 
\hline
\multicolumn{3}{|c|}{\bf Pion masses for $\beta=2.05$.}\\ \hline
\end{tabular}
& \hspace{0.5in} &
\begin{tabular}{|ccc|} \hline
 m & pion & error \\ \hline 
-1.00 & 1.569 & 0.014 \\ 
-1.05 & 1.488 & 0.012 \\ 
-1.10 & 1.385 & 0.025 \\ 
-1.15 & 1.254 & 0.022 \\ 
-1.20 & 1.062 & 0.038 \\ 
-1.25 & 0.592 & 0.043 \\ 
-1.30 & 0.413 & 0.018 \\ 
-1.35 & 0.915 & 0.021 \\ 
-1.40 & 1.228 & 0.016 \\ 
-1.45 & 1.386 & 0.028 \\ 
-1.50 & 1.419 & 0.017 \\ 
-1.55 & 1.441 & 0.017 \\ 
\hline
\multicolumn{3}{|c|}{\bf Pion masses for $\beta=2.10$.}\\ \hline
\end{tabular}
\end{tabular}
\end{center}

\begin{center}
\begin{tabular}{ccc}
\begin{tabular}{|ccc|} \hline
 m & pion & error \\ \hline 
0.50 & 2.776 & 0.026 \\ 
0.25 & 2.632 & 0.031 \\ 
0.00 & 2.489 & 0.024 \\ 
-0.25 & 2.328 & 0.027 \\ 
-0.50 & 2.144 & 0.030 \\ 
-0.75 & 1.923 & 0.032 \\ 
-1.00 & 1.662 & 0.030 \\ 
-1.10 & 1.540 & 0.027 \\ 
-1.15 & 1.445 & 0.028 \\ 
-1.20 & 1.363 & 0.029 \\ 
-1.25 & 1.243 & 0.019 \\ 
-1.30 & 1.077 & 0.046 \\ 
-1.35 & 0.327 & 0.035 \\ 
-1.40 & 0.969 & 0.039 \\ 
-1.45 & 1.247 & 0.030 \\ 
-1.50 & 1.378 & 0.033 \\ 
-1.55 & 1.399 & 0.025 \\ 
-1.60 & 1.383 & 0.019 \\ 
-1.65 & 1.394 & 0.041 \\ 
-1.75 & 1.315 & 0.050 \\ 
-2.00 & 0.936 & 0.043 \\ 
\hline
\multicolumn{3}{|c|}{\bf Pion masses for $\beta=2.00$.}\\ \hline
\end{tabular}
& \hspace{0.5in} &
\begin{tabular}{|ccc|} \hline
 m & pion & error \\ \hline 
0.50 & 2.735 & 0.025 \\ 
0.25 & 2.582 & 0.023 \\ 
0.00 & 2.448 & 0.039 \\ 
-0.25 & 2.254 & 0.030 \\ 
-0.50 & 2.013 & 0.035 \\ 
-0.75 & 1.733 & 0.041 \\ 
-0.90 & 1.488 & 0.018 \\ 
-0.95 & 1.392 & 0.019 \\ 
-1.00 & 1.265 & 0.022 \\ 
-1.05 & 1.112 & 0.026 \\ 
-1.10 & 0.889 & 0.032 \\ 
-1.15 & 0.571 & 0.026 \\ 
-1.20 & 0.245 & 0.0058 \\ 
-1.25 & 0.5721 & 0.0054 \\ 
-1.30 & 1.018 & 0.017 \\ 
-1.35 & 1.283 & 0.020 \\ 
-1.40 & 1.421 & 0.021 \\ 
-1.45 & 1.476 & 0.025 \\ 
-1.50 & 1.494 & 0.026 \\ 
-1.75 & 1.309 & 0.052 \\ 
-2.00 & 0.903 & 0.072 \\ 
\hline
\multicolumn{3}{|c|}{\bf Pion masses for $\beta=2.25$.}\\ \hline
\end{tabular}
\end{tabular}
\end{center}

\begin{center}
\begin{tabular}{ccc}
\begin{tabular}{|ccc|} \hline
 m & pion & error \\ \hline 
-0.80 & 1.520 & 0.021 \\ 
-0.85 & 1.414 & 0.030 \\ 
-0.90 & 1.303 & 0.015 \\ 
-0.95 & 1.171 & 0.022 \\ 
-1.00 & 0.989 & 0.016 \\ 
-1.05 & 0.787 & 0.026 \\ 
-1.10 & 0.497 & 0.015 \\ 
-1.15 & 0.2383 & 0.0085 \\ 
-1.20 & 0.499 & 0.011 \\ 
-1.25 & 0.882 & 0.032 \\ 
-1.30 & 1.193 & 0.026 \\ 
-1.35 & 1.388 & 0.019 \\ 
\hline
\multicolumn{3}{|c|}{\bf Pion masses for $\beta=2.35$.}\\ \hline
\end{tabular}
& \hspace{0.5in} &
\begin{tabular}{|ccc|} \hline
 m & pion & error \\ \hline 
-0.75 & 1.525 & 0.023 \\ 
-0.80 & 1.441 & 0.039 \\ 
-0.85 & 1.322 & 0.026 \\ 
-0.90 & 1.187 & 0.047 \\ 
-0.95 & 1.049 & 0.029 \\ 
-1.00 & 0.859 & 0.047 \\ 
-1.05 & 0.623 & 0.030\\ 
-1.10 & 0.319 & 0.012 \\ 
-1.15 & 0.310 & 0.011 \\ 
-1.20 & 0.634 & 0.024 \\ 
-1.25 & 0.998 & 0.029 \\ 
-1.30 & 1.286 & 0.031 \\ 
\hline
\multicolumn{3}{|c|}{\bf Pion masses for $\beta=2.40$.}\\ \hline
\end{tabular}
\end{tabular}
\end{center}

\begin{center}
\begin{tabular}{ccc}
\begin{tabular}{|ccc|} \hline
 m & pion & error \\ \hline 
0.50 & 2.646 & 0.026 \\ 
0.25 & 2.474 & 0.023 \\ 
0.00 & 2.302 & 0.031\\ 
-0.25 & 2.063 & 0.037 \\ 
-0.50 & 1.809 & 0.025 \\ 
-0.70 & 1.490 & 0.012 \\ 
-0.75 & 1.388 & 0.016 \\ 
-0.80 & 1.287 & 0.018 \\ 
-0.85 & 1.136 & 0.016 \\ 
-0.90 & 0.992 & 0.017 \\ 
-0.95 & 0.814 & 0.013 \\ 
-1.00 & 0.585 & 0.012 \\ 
-1.05 & 0.3278 & 0.0089 \\ 
-1.10 & 0.2660 & 0.0069\\ 
-1.15 & 0.5422 & 0.0066 \\ 
-1.20 & 0.895 & 0.027 \\ 
-1.25 & 1.205 & 0.022 \\ 
-1.50 & 1.524 & 0.048 \\ 
-1.75 & 1.252 & 0.094 \\ 
-2.00 & 0.820 & 0.040 \\ 
\hline
\multicolumn{3}{|c|}{\bf Pion masses for $\beta=2.50$.}\\ \hline
\end{tabular}
& \hspace{0.5in} &
\begin{tabular}{|ccc|} \hline
 m & pion & error \\ \hline 
0.50 & 2.556 & 0.030 \\ 
0.25 & 2.383 & 0.037 \\ 
0.00 & 2.177 & 0.031 \\ 
-0.25 & 1.916 & 0.034 \\ 
-0.50 & 1.569 & 0.040 \\ 
-0.65 & 1.252 & 0.028 \\ 
-0.70 & 1.109 & 0.039 \\ 
-0.75 & 1.002 & 0.049 \\ 
-0.80 & 0.860 & 0.033 \\ 
-0.85 & 0.686 & 0.033 \\ 
-0.90 & 0.494 & 0.020 \\ 
-0.95 & 0.3026 & 0.0092 \\ 
-1.00 & 0.243 & 0.013 \\ 
-1.05 & 0.428 & 0.017 \\ 
-1.10 & 0.736 & 0.020 \\ 
-1.15 & 1.040 & 0.040 \\ 
-1.20 & 1.273 & 0.044 \\ 
-1.25 & 1.448 & 0.051 \\ 
-1.30 & 1.563 & 0.036 \\ 
-1.35 & 1.579 & 0.042 \\ 
-1.50 & 1.529 & 0.035 \\ 
-1.75 & 1.260 & 0.061 \\ 
-2.00 & 0.777 & 0.032 \\ 
\hline
\multicolumn{3}{|c|}{\bf Pion masses for $\beta=2.75$.}\\ \hline
\end{tabular}
\end{tabular}
\end{center}

\begin{center}
\begin{tabular}{|ccc|} \hline
 m & pion & error \\ \hline 
0.50 & 2.482 & 0.018 \\ 
0.25 & 2.275 & 0.023 \\ 
0.00 & 2.019 & 0.030 \\ 
-0.25 & 1.731 & 0.041 \\ 
-0.50 & 1.340 & 0.050 \\ 
-0.60 & 1.064 & 0.045 \\ 
-0.65 & 0.941 & 0.045 \\ 
-0.70 & 0.799 & 0.050 \\ 
-0.75 & 0.641 & 0.022 \\ 
-0.80 & 0.496 & 0.033 \\ 
-0.85 & 0.330 & 0.014 \\ 
-0.90 & 0.231 & 0.010 \\ 
-1.00 & 0.327 & 0.020 \\ 
-1.05 & 0.541 & 0.013 \\ 
-1.10 & 0.823 & 0.039 \\ 
-1.15 & 1.100 & 0.047 \\ 
-1.20 & 1.357 & 0.042 \\ 
-1.25 & 1.488 & 0.058 \\ 
-1.30 & 1.572 & 0.021 \\ 
-1.50 & 1.441 & 0.35 \\ 
-1.75 & 1.279 & 0.045 \\ 
-2.00 & 0.690 & 0.024 \\ 
\hline
\multicolumn{3}{|c|}{\bf Pion masses for $\beta=3.00$.}\\ \hline
\end{tabular}
\end{center}

\newpage

\section*{Appendix B:  Rho masses}

\begin{center}
\begin{tabular}{ccc}
\begin{tabular}{|ccc|} \hline
 m & rho & error \\ \hline 
0.50 & 2.830 & 0.011 \\ 
0.25 & 2.718 & 0.014 \\ 
0.00 & 2.5605 & 0.0070 \\ 
-0.25 & 2.418 & 0.013 \\ 
-0.50 & 2.252 & 0.010 \\ 
-0.75 & 2.0551 & 0.0072 \\ 
-1.00 & 1.841 & 0.019 \\ 
-1.25 & 1.622 & 0.024 \\ 
-1.50 & 1.323 & 0.016 \\ 
-1.65 & 1.110 & 0.010 \\ 
-1.70 & 1.506 & 0.013 \\ 
-1.75 & 1.499 & 0.013 \\ 
-1.80 & 1.463 & 0.015 \\ 
-1.85 & 1.420 & 0.010 \\ 
-1.90 & 1.347 & 0.018 \\ 
\hline
\multicolumn{3}{|c|}{\bf Rho masses for $\beta=1.50$.}\\ \hline
\end{tabular}
& \hspace{0.5in} &
\begin{tabular}{|ccc|} \hline
 m & rho & error \\ \hline 
0.50 & 2.8119 & 0.0074 \\ 
0.25 & 2.7012 & 0.0080\\ 
0.00 & 2.5366 & 0.0093 \\ 
-0.25 & 2.387 & 0.013 \\ 
-0.50 & 2.221 & 0.025 \\ 
-0.75 & 2.029 & 0.014 \\ 
-1.00 & 1.796 & 0.016 \\ 
-1.25 & 1.557 & 0.017 \\ 
-1.40 & 1.3633 & 0.0051 \\ 
-1.45 & 1.302 & 0.012 \\ 
-1.50 & 1.221 & 0.0099 \\ 
-1.55 & 1.149 & 0.0078 \\ 
-1.60 & 1.389 & 0.28 \\ 
-1.65 & 1.567 & 0.015 \\ 
-1.70 & 1.530 & 0.014 \\ 
-1.75 & 1.4803 & 0.0092 \\ 
\hline
\multicolumn{3}{|c|}{\bf Rho masses for $\beta=1.75$.}\\ \hline
\end{tabular}
\end{tabular}
\end{center}

\begin{center}
\begin{tabular}{ccc}
\begin{tabular}{|ccc|} \hline
 m & rho & error \\ \hline 
-1.20 & 1.534 & 0.0073 \\ 
-1.25 & 1.474 & 0.020 \\ 
-1.30 & 1.428 & 0.022 \\ 
-1.35 & 1.344 & 0.018 \\ 
-1.40 & 1.215 & 0.017 \\ 
-1.45 & 1.247 & 0.058 \\ 
-1.50 & 1.560 & 0.045 \\ 
-1.55 & 1.620 & 0.011 \\ 
-1.60 & 1.596 & 0.020 \\ 
-1.65 & 1.569 & 0.018 \\ 
-1.70 & 1.526 & 0.032 \\ 
-1.75 & 1.474 & 0.014 \\ 
\hline
\multicolumn{3}{|c|}{\bf Rho masses for $\beta=1.90$.}\\ \hline
\end{tabular}
& \hspace{0.5in}
\begin{tabular}{|ccc|} \hline
 m & rho & error \\ \hline 
-1.15 & 1.5674 & 0.0059 \\ 
-1.20 & 1.513 & 0.012 \\ 
-1.25 & 1.439 & 0.011 \\ 
-1.30 & 1.351 & 0.014 \\ 
-1.35 & 1.2269 & 0.0099 \\ 
-1.40 & 0.65 & 0.41 \\ 
-1.45 & 1.408 & 0.018 \\ 
-1.50 & 1.6033 & 0.0091 \\ 
-1.55 & 1.635 & 0.010 \\ 
-1.60 & 1.6084 & 0.0078 \\ 
-1.65 & 1.564 & 0.013 \\ 
-1.70 & 1.527 & 0.012 \\ 
\hline
\multicolumn{3}{|c|}{\bf Rho masses for $\beta=1.95$.}\\ \hline
\end{tabular}
\end{tabular}
\end{center}

\begin{center}
\begin{tabular}{ccc}
\begin{tabular}{|ccc|} \hline
 m & rho & error \\ \hline 
-1.05 & 1.6170 & 0.0077 \\ 
-1.10 & 1.542 & 0.011 \\ 
-1.15 & 1.4678 & 0.0046 \\ 
-1.20 & 1.344 & 0.012 \\ 
-1.25 & 1.157 & 0.012 \\ 
-1.30 & 0.470 & 0.048 \\ 
-1.35 & 0.754 & 0.011 \\ 
-1.40 & 1.329 & 0.012 \\ 
-1.45 & 1.597 & 0.013 \\ 
-1.50 & 1.6527 & 0.0066 \\ 
-1.55 & 1.627 & 0.010 \\ 
-1.60 & 1.604 & 0.012 \\ 
\hline
\multicolumn{3}{|c|}{\bf Rho masses for $\beta=2.05$.}\\ \hline
\end{tabular}
& \hspace{0.5in} &
\begin{tabular}{|ccc|} \hline
 m & rho & error \\ \hline 
-1.00 & 1.6278 & 0.0044 \\ 
-1.05 & 1.5596 & 0.0050 \\ 
-1.10 & 1.461 & 0.011 \\ 
-1.15 & 1.346 & 0.012 \\ 
-1.20 & 1.162 & 0.015 \\ 
-1.25 & 0.645 & 0.031 \\ 
-1.30 & 0.424 & 0.010 \\ 
-1.35 & 1.042 & 0.029 \\ 
-1.40 & 1.4635 & 0.0093 \\ 
-1.45 & 1.647 & 0.016 \\ 
-1.50 & 1.644 & 0.011 \\ 
-1.55 & 1.626 & 0.011 \\ 
\hline
\multicolumn{3}{|c|}{\bf Rho masses for $\beta=2.10$.}\\ \hline
\end{tabular}
\end{tabular}
\end{center}

\begin{center}
\begin{tabular}{ccc}
\begin{tabular}{|ccc|} \hline
 m & rho & error \\ \hline 
0.50 & 2.786 & 0.014 \\ 
0.25 & 2.647 & 0.019 \\ 
0.00 & 2.501 & 0.017 \\ 
-0.25 & 2.341 & 0.013 \\ 
-0.50 & 2.174 & 0.017 \\ 
-0.75 & 1.960 & 0.015 \\ 
-1.00 & 1.714 & 0.014 \\ 
-1.10 & 1.604 & 0.014 \\ 
-1.15 & 1.519 & 0.011 \\ 
-1.20 & 1.449 & 0.016 \\ 
-1.25 & 1.353 & 0.012 \\ 
-1.30 & 1.201 & 0.0081 \\ 
-1.35 & 0.384 & 0.018 \\ 
-1.40 & 1.142 & 0.032 \\ 
-1.45 & 1.519 & 0.029 \\ 
-1.50 & 1.635 & 0.014 \\ 
-1.55 & 1.630 & 0.013 \\ 
-1.60 & 1.590 & 0.011 \\ 
-1.65 & 1.565 & 0.015 \\ 
-1.75 & 1.449 & 0.022 \\ 
-2.00 & 0.995 & 0.033 \\ 
\hline
\multicolumn{3}{|c|}{\bf Rho masses for $\beta=2.00$.}\\ \hline
\end{tabular}
& \hspace{0.5in} &
\begin{tabular}{|ccc|} \hline
 m & rho & error \\ \hline 
0.50 & 2.739 & 0.012 \\ 
0.25 & 2.597 & 0.014 \\ 
0.00 & 2.456 & 0.025 \\ 
-0.25 & 2.276 & 0.015 \\ 
-0.50 & 2.038 & 0.013 \\ 
-0.75 & 1.774 & 0.027 \\ 
-0.90 & 1.538 & 0.013 \\ 
-0.95 & 1.452 & 0.012 \\ 
-1.00 & 1.331 & 0.010 \\ 
-1.05 & 1.1794 & 0.0080 \\ 
-1.10 & 0.953 & 0.023 \\ 
-1.15 & 0.606 & 0.021 \\ 
-1.20 & 0.2456 & 0.0035 \\ 
-1.25 & 0.5973 & 0.0058 \\ 
-1.30 & 1.132 & 0.012 \\ 
-1.35 & 1.497 & 0.018 \\ 
-1.40 & 1.654 & 0.012 \\ 
-1.45 & 1.683 & 0.010 \\ 
-1.50 & 1.671 & 0.021 \\ 
-1.75 & 1.420 & 0.032 \\ 
-2.00 & 0.919 & 0.054 \\ 
\hline
\multicolumn{3}{|c|}{\bf Rho masses for $\beta=2.25$.}\\ \hline
\end{tabular}
\end{tabular}
\end{center}

\begin{center}
\begin{tabular}{ccc}
\begin{tabular}{|ccc|} \hline
 m & rho & error \\ \hline 
-0.80 & 1.5653 & 0.0088 \\ 
-0.85 & 1.459 & 0.017 \\ 
-0.90 & 1.3524 & 0.0067 \\ 
-0.95 & 1.226 & 0.011 \\ 
-1.00 & 1.039 & 0.016 \\ 
-1.05 & 0.830 & 0.028 \\ 
-1.10 & 0.519 & 0.017 \\ 
-1.15 & 0.230 & 0.0058 \\ 
-1.20 & 0.5212 & 0.0080 \\ 
-1.25 & 0.951 & 0.017 \\ 
-1.30 & 1.364 & 0.029 \\ 
-1.35 & 1.612 & 0.013 \\ 
\hline
\multicolumn{3}{|c|}{\bf Rho masses for $\beta=2.35$.}\\ \hline
\end{tabular}
& \hspace{0.5in} &
\begin{tabular}{|ccc|} \hline
 m & rho & error \\ \hline 
-0.75 & 1.566 & 0.013 \\ 
-0.80 & 1.481 & 0.018 \\ 
-0.85 & 1.366 & 0.016 \\ 
-0.90 & 1.228 & 0.019 \\ 
-0.95 & 1.100 & 0.029 \\ 
-1.00 & 0.895 & 0.040 \\ 
-1.05 & 0.645 & 0.016 \\ 
-1.10 & 0.320 & 0.013 \\ 
-1.15 & 0.3028 & 0.0058 \\ 
-1.20 & 0.6610 & 0.0094 \\ 
-1.25 & 1.099 & 0.019 \\ 
-1.30 & 1.471 & 0.020 \\ 
\hline
\multicolumn{3}{|c|}{\bf Rho masses for $\beta=2.40$.}\\ \hline
\end{tabular}
\end{tabular}
\end{center}

\begin{center}
\begin{tabular}{ccc}
\begin{tabular}{|ccc|} \hline
 m & rho & error \\ \hline 
0.50 & 2.652 & 0.014 \\ 
0.25 & 2.4906 & 0.0077 \\ 
0.00 & 2.315 & 0.017 \\ 
-0.25 & 2.085 & 0.011 \\ 
-0.50 & 1.8351 & 0.0068 \\ 
-0.70 & 1.5231 & 0.0075 \\ 
-0.75 & 1.422 & 0.010 \\ 
-0.80 & 1.3252 & 0.0094 \\ 
-0.85 & 1.1699 & 0.0097 \\ 
-0.90 & 1.030 & 0.012 \\ 
-0.95 & 0.8384 & 0.0059 \\ 
-1.00 & 0.605 & 0.012 \\ 
-1.05 & 0.3239 & 0.0055 \\ 
-1.10 & 0.2584 & 0.0033 \\ 
-1.15 & 0.5430 & 0.0067 \\ 
-1.20 & 0.952 & 0.012 \\ 
-1.25 & 1.331 & 0.013 \\ 
-1.50 & 1.647 & 0.024 \\ 
-1.75 & 1.345 & 0.040 \\ 
-2.00 & 0.807 & 0.023 \\ 
\hline
\multicolumn{3}{|c|}{\bf Rho masses for $\beta=2.50$.}\\ \hline
\end{tabular}
& \hspace{0.5in} &
\begin{tabular}{|ccc|} \hline
 m & rho & error \\ \hline 
0.50 & 2.563 & 0.0057 \\ 
0.25 & 2.390 & 0.019 \\ 
0.00 & 2.190 & 0.012 \\ 
-0.25 & 1.926 & 0.025 \\ 
-0.50 & 1.597 & 0.018 \\ 
-0.65 & 1.282 & 0.027 \\ 
-0.70 & 1.121 & 0.028 \\ 
-0.75 & 1.016 & 0.031 \\ 
-0.80 & 0.900 & 0.019 \\ 
-0.85 & 0.689 & 0.030 \\ 
-0.90 & 0.4899 & 0.0078 \\ 
-0.95 & 0.2935 & 0.0099 \\ 
-1.00 & 0.2271 & 0.0073 \\ 
-1.05 & 0.4207 & 0.0071 \\ 
-1.10 & 0.742 & 0.011 \\ 
-1.15 & 1.088 & 0.041 \\ 
-1.20 & 1.417 & 0.033 \\ 
-1.25 & 1.638 & 0.030 \\ 
-1.30 & 1.740 & 0.019 \\ 
-1.35 & 1.736 & 0.023 \\ 
-1.50 & 1.636 & 0.024 \\ 
-1.75 & 1.310 & 0.026 \\ 
-2.00 & 0.761 & 0.018 \\ 
\hline
\multicolumn{3}{|c|}{\bf Rho masses for $\beta=2.75$.}\\ \hline
\end{tabular}
\end{tabular}
\end{center}

\begin{center}
\begin{tabular}{|ccc|} \hline
 m & rho & error \\ \hline 
0.50 & 2.490 & 0.010 \\ 
0.25 & 2.281 & 0.011 \\ 
0.00 & 2.028 & 0.015 \\ 
-0.25 & 1.750 & 0.017 \\ 
-0.50 & 1.358 & 0.024 \\ 
-0.60 & 1.069 & 0.028 \\ 
-0.65 & 0.953 & 0.016 \\ 
-0.70 & 0.807 & 0.032 \\ 
-0.75 & 0.629 & 0.011 \\ 
-0.80 & 0.488 & 0.015 \\ 
-0.85 & 0.317 & 0.012 \\ 
-0.90 & 0.2123 & 0.0060 \\ 
-0.95 & 0.305 & 0.011 \\ 
-1.00 & 0.527 & 0.013 \\ 
-1.05 & 0.835 & 0.018 \\ 
-1.10 & 1.147 & 0.032 \\ 
-1.15 & 1.465 & 0.033 \\ 
-1.20 & 1.652 & 0.033 \\ 
-1.25 & 1.738 & 0.025 \\ 
-1.30 & 1.764 & 0.020 \\ 
-1.50 & 1.621 & 0.022 \\ 
-1.75 & 1.301 & 0.031 \\ 
-2.00 & 0.681 & 0.015 \\ 
\hline
\multicolumn{3}{|c|}{\bf Rho masses for $\beta=3.00$.}\\ \hline
\end{tabular}
\end{center}

\end{document}